\documentclass[prl,twocolumn,showpacs,superscriptaddress]{revtex4}

\usepackage{graphicx}

\usepackage{ulem}
\usepackage{color}
\usepackage{amsmath,amssymb}

\def\Ueff{U_\mathrm{eff}}
\def\TMIT{T_\mathrm{MIT}}
\def\Asia{A_\mathrm{sia}}
\def\ADM{A_\mathrm{DM}}
\def\Dd{\Delta_\mathrm{D}}
\def\Dc{\Delta_\mathrm{C}}
\def\mos{m_\mathrm{Os}}

\begin{document}
\title{Noncollinear magnetism and spin-orbit coupling in 5$d$ pyrochlore oxide Cd$_2$Os$_2$O$_7$}
\author{Hiroshi Shinaoka}
\affiliation{Nanosystem Research Institute ``RICS'', National Institute of Advanced Industrial Science and Technology (AIST), Umezono, Tsukuba 305-8568, Japan}
\affiliation{Japan Science and Technology Agency (JST), CREST, Honcho, Kawaguchi, Saitama 332-0012, Japan}
\author{Takashi Miyake}
\affiliation{Nanosystem Research Institute ``RICS'', National Institute of Advanced Industrial Science and Technology (AIST), Umezono, Tsukuba 305-8568, Japan}
\affiliation{Japan Science and Technology Agency (JST), CREST, Honcho, Kawaguchi, Saitama 332-0012, Japan}
\author{Shoji Ishibashi}
\affiliation{Nanosystem Research Institute ``RICS'', National Institute of Advanced Industrial Science and Technology (AIST), Umezono, Tsukuba 305-8568, Japan}
\affiliation{Japan Science and Technology Agency (JST), CREST, Honcho, Kawaguchi, Saitama 332-0012, Japan}
\date{\today}

\begin{abstract}
We investigated the electronic and magnetic properties of the pyrochlore oxide Cd$_2$Os$_2$O$_7$ using the density-functional theory plus on-site repulsion ($U$) method, and depict the ground-state phase diagram with respect to $U$.
We conclude that the all-in/all-out non-collinear magnetic order is stable in a wide range of $U$.
We also show that the easy-axis anisotropy arising from the spin-orbit (SO) coupling plays a significant role in stabilizing the all-in/all-out magnetic order.
A \textit{pseudo gap} was observed near the transition between the antiferromagnetic metallic and insulating phases.
Finally, we discuss possible origins of the peculiar low-temperature($T$) properties observed in experiments.
\end{abstract}


\maketitle

Pyrochlore transition-metal oxides $A_2B_2$O$_7$ are a class of materials that have been extensively investigated for several decades~\cite{Gardner10}.
Recently, particular attention has been paid to 5$d$ transition-metal oxides, such as iridates ($A_2$Ir$_2$O$_7$),
in search for unconventional phenomena that are induced by the competing spin-orbit (SO) coupling and electron correlation.

Cd$_2$Os$_2$O$_7$ is one of the few compounds whose low-$T$ magnetic structure has been experimentally determined 
among a number of magnetic 5$d$ pyrochlore oxides.
The Cd$^{2+}$ ion is non-magnetic, whereas the Os$^{5+}$ ion with a 5$d^3$ configuration can be magnetic.
This compound exhibits a purely electronic continuous metal-insulator transition (MIT) 
concurrently with a N\'{e}el ordering at $T_\mathrm{MIT}\simeq 227$ K~\cite{Sleight74,Mandrus01,Matsuda11,Yamaura11}.
Magnetic-susceptibility~\cite{Sleight74,Mandrus01} and $\mu$SR measurements~\cite{Koda07} suggested the N\'{e}el ordering,
whereas powder neutron diffraction did not confirm the abovementioned observation~\cite{Reading01}.
Recently, Yamaura \textit{et al.} have successfully detected a magnetic reflection at the wave vector of $\boldsymbol{q}=0$
using resonant X-ray scattering on high-quality single crystals~\cite{Yamaura11}.
They showed that only the so-called all-in/all-out magnetic order [see Fig.~1(a)] is compatible with the cubic symmetry of the crystal among the possible $\boldsymbol{q}=0$ magnetic orders.
The all-in/all-out magnetic ordering has been suggested also in other 5$d$ pyrochlore oxides, i.e, Nd$_2$Ir$_2$O$_7$ in experimental work~\cite{Tomiyasu11} and Y$_2$Ir$_2$O$_7$ in theoretical work~\cite{Wan11}.

Despite the recent experimental progress, we are still far from fully understanding the electronic properties of this compound.
In particular, several puzzling electronic properties have been reported experimentally: 
(1) In contrast to the opening of an optical gap of the order of $800~\mathrm{cm}^{-1}$ ($\simeq 1100~\mathrm{K}$)~\cite{Padilla02},
the semiconducting gap continuously vanishes toward low $T$'s~\cite{Sleight74, Mandrus01}.
This indicates the absence of a clear charge gap at low $T$'s which seemingly contradicts the semiconducting behavior of the resistivity up to $\TMIT$.
(2) The N\'{e}el transition temperature of this compound ($=T_\mathrm{MIT}$) is, to the beset of our knowledge, one of the highest among the magnetic pyrochlore oxides~\cite{Gardner10}.
This observation is quite surprising because geometrical frustration tends to prevent long-range magnetic ordering.
These puzzling characteristics and the similarity with the Ir oxides urged us to investigate this compound as a prototype of 5$d$ pyrochlore oxides.

\begin{figure}[!]
 \centering
 \includegraphics[width=.475\textwidth,clip]{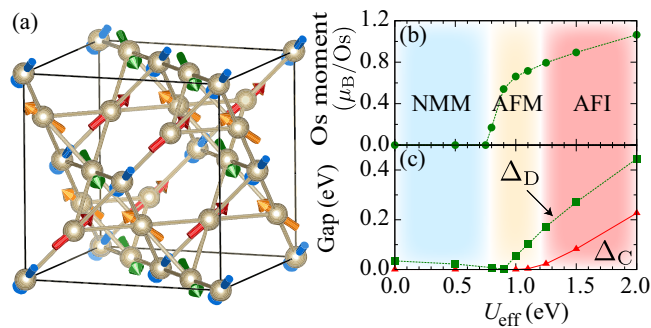}
 \caption{(color online). (a) Cubic unit cell of Cd$_2$Os$_2$O$_7$ contains 8 formula units.
Only the corner sharing tetrahedron network of Os atoms is shown for clarity.
The arrows represent the all-in/all-out magnetic order with the ordering vector of $\boldsymbol{q}=0$.
 (b)/(c) $\Ueff$ dependences of the Os magnetic moment $\mos$, the charge gap $\Dc$, and the direct gap $\Dd$.
 The phase diagram consists of a non-magnetic metal (NMM), an antiferromagnetic metal (AFM), and an antiferromagnetic insulator (AFI).
}
 \label{fig:pd}
\end{figure}

In this Letter, we investigate the ground-state properties of Cd$_2$Os$_2$O$_7$ using extensive LSDA+SO+$U$ calculations (LSDA denotes the local spin density approximation).
We explore the ground-state phase diagram as a function of $\Ueff~(\equiv U-J)$.
Here, $\Ueff$ is an empirical parameter, and $\Ueff = 1$--$2$ eV is expected to be appropriate for spatially extended $5d$ orbitals~\cite{Wan11}.
We show that the all-in/all-out magnetic order is stable in an antiferromagnetic metal (AFM) and in an antiferromagnetic insulator (AFI).
We observe that strong magnetic anisotropy arising from the SO coupling stabilizes the all-in/all-out magnetic order.
Finally, we discuss possible origins of the experimentally-observed puzzling characteristics of Cd$_2$Os$_2$O$_7$.

In the following calculations, we use a fully-relativistic two-component first-principles computational code, QMAS (Quantum MAterials Simulator)~\cite{qmas}.
We employ the projector augmented-wave method~\cite{Blochl94b} and the LSDA+SO+$U$ method~\cite{Ceperley80,Perdew81,Dudarev98,Oda98,Kosugi_Au}.
The relativistic effect including the SO coupling is fully considered in solving the relativistic Kohn-Sham equation~\cite{Oda98,Kosugi_Au}.
In the following calculations, we adopt a face-centered cubic primitive unit cell for Cd$_2$Os$_2$O$_7$ containing two formula units,
and restrict consideration to $\boldsymbol{q}=0$ magnetic ordering.
Brillouin-zone integrations were performed using up to $12\times 12\times 12$ $k$-point samplings using the improved tetrahedron method~\cite{Blochl94}.
We used a planewave cutoff energy of 40 Ry.
The following calculations were done with the experimental lattice structure at 180 K: $a=10.1598$ \AA~and $x(\mathrm{O}_1)=0.319$~\cite{Mandrus01}.
Every Os atom is located at the center of an OsO$_6$ octahedron.
For $x(\mathrm{O}_1)>0.3125$, each oxygen octahedron is slightly compressed along the local $\langle 111 \rangle$ axis that connects the centers of the two neighboring Os tetrahedra.

To identify the antiferromagnetic ordering, we calculate the local magnetic moment projected on an Os atom, $\mos$,
by integrating the magnetic moment within a radius of 2.5 a.u. ($=1.323$ \AA).
The direct gap, $\Dd$, is defined as the minimum gap between the conduction and valence bands,
which approximately corresponds to the optical gap.
In the following discussion, $\Dc$ denotes the charge gap, which identifies the MIT.

\begin{figure}[t]
 \centering
 \includegraphics[width=.49\textwidth,clip]{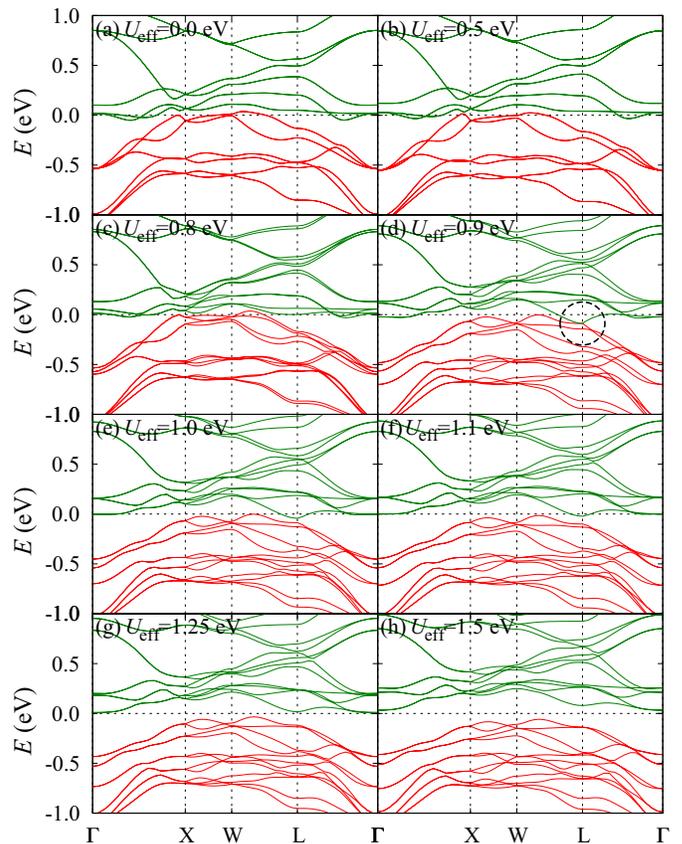}
 \caption{(color online). Band structures on high-symmetry lines computed at $\Ueff=0.0$, 0.5, 0.8, 0.9, 1.0, 1.1, 1.25, and 1.5 eV.
 Energy, $E$, is measured from the Fermi level ($E_\mathrm{F}$) [(a)-(f)] or from the center of the charge gap [(g), (h)].
 Conduction and valence bands are denoted by green and red lines, respectively. The direct gap closes near the $L$ point at $\Ueff\simeq 0.9$ eV [see the circle in (d)].
 }
 \label{fig:2}
\end{figure}
\begin{figure}[t]
 \centering
 \includegraphics[width=.475\textwidth,clip]{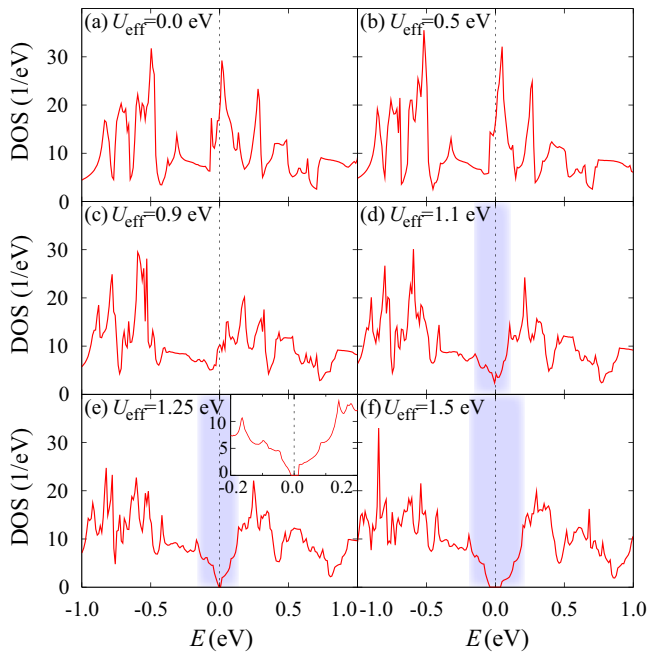}
 \caption{(color online). Density of states computed at $\Ueff=0.0$, 0.5, 0.9, 1.1, 1.25, and 1.5 eV for the same data as in Fig.~2.
 The inset of (e) is an enlarged plot of (e) near the charge gap. The density of states is measured per primitive unit cell.
 The shaded areas denote a \textit{pseudo gap}, which appears near the MIT.}
 \label{fig:3}
\end{figure}

Figure~1(b) shows the computed ground-state phase diagram with respect to $\Ueff$.
At small $\Ueff$'s, the ground state is non-magnetic metal (NMM).
By increasing $\Ueff$,
the ground state turns into the AFM phase at $\Ueff\simeq 0.75$ eV, and further into the AFI phase at $\Ueff\simeq 1.2$ eV.
We found that the all-in/all-out magnetic order is the most stable in the entire parameter region of the AFM and AFI phases:
Os moments with the same magnitude point toward or away from the centers of the tetrahedra along the local $\langle 111\rangle$ axes as illustrated in Fig.~1(a).
We obtained $\mos\simeq 0.8$--$1.1$ $\mu_\mathrm{B}/\mathrm{Os}$ in the AFI phase, which is considerably smaller than 3~$\mu_\mathrm{B}$/Os for the high-spin state.

Figures~2(a)-(b) show the calculated electronic band structures near the Fermi level.
At $\Ueff=0.0$ eV [Fig.~2(a)], the Fermi level lies in the half-filled $t_\mathrm{2g}$ bands consisting of twelve Kramers(doubly)-degenerate bands.
Note that the effect of the distortion of the octahedral crystal field is not strong enough to split the $t_\mathrm{2g}$ manifold.
As suggested by the previous LSDA studies~\cite{Singh02,Harima02},
the band structure looks semi-metallic on the high-symmetry lines.
In fact, we obtained $\Dd\simeq 0.035$ eV at $\Ueff=0.0$ eV.
Although $\Dd$ decreases with increasing $\Ueff$, $\Dd$ remains non zero within the NMM phase as shown in Fig.~1(c).
Figures 3(a)-(f) show the calculated electronic density of states.
At $\Ueff=0.0$ eV [Fig.~3(a)], one clearly see a sharp peak near the Fermi level, which may originate in low-lying flat bands.
The low-energy structure of the density of states remains essentially unchanged within the NMM phase [see Fig.~3(b)].

With increasing $\Ueff$, the ground state turns into the AFM phase at $\Ueff\simeq 0.8$ eV.
We found that the all-in/all-out magnetic order is the most stable in the AFM and AFI phases.
We used various initial spin and charge densities for the iterative scheme,
but we could not find any other stable magnetic solutions.
In the AFM phase, the time-reversal symmetry breaking lifts the Kramers degeneracy in the NM band structure.
Note that the AF transition is associated with no Brillouin-zone folding because the all-in/all-out order preserves the translational symmetry of the lattice.
The direct gap, $\Dd$, closes at $\Ueff\simeq 0.9$ eV
with a band inversion between the valence and conduction bands [see Fig.~1(c) and the circle in Fig.~2(d)]
\footnote{The product of the parity eigenvalues for the occupied bands changes its sign between $\Ueff=0.8$ and 0.9 eV at the $L$ point.}.
As $\mos$ develops with increasing $\Ueff$, $\Dd$ opens again and becomes increasingly larger,
finally resulting in the continuous MIT at $\Ueff\simeq 1.2$ eV [Fig.~1(c)].
The MIT is characterized by vanishing electron and hole Fermi surfaces (Lifshitz transition).
This is clearly distinguished from the Slater transition in which a charge gap appears because of a magnetic superlattice structure~\cite{Slater51}.
Another notable observation near the MIT is that the density of states is considerably suppressed near the Fermi level up to about 0.2 eV, which is remarkably higher than $\Dc$ [Figs.~3(d)-(f)].
As seen in Fig.~3(d), this \textit{pseudo gap} starts to develop in the AFM phase,
suggesting that it comes from the modification of the band structure near the Fermi level with the emergent magnetic order.

Next, we discuss the magnetic properties of the AFI phase.
It is well known that pyrochlore antiferromagnets tend to be frustrated when the spins are isotropic.
This is clearly seen in the nearest-neighbor classical antiferromagnet which exhibits no phase transition and remains paramagnetic down to zero $T$ with a macroscopic degeneracy~\cite{Moessner98}.
The ground-state manifold consists of degenerate states
in each of which the summation of four spin moments vanishes on every tetrahedron. 
The degeneracy can be lifted by magnetic anisotropy which arises from the SO coupling.
In particular, the all-in/all-out order is selected as the unique ground state by the local $\langle 111 \rangle$ easy-axis anisotropy~\cite{Bramwell94,Bramwell98,Moessner98b}.

Figure~4(a) shows magnetic anisotropy energies calculated for $\Ueff=1.25$ and $2.0$ eV, respectively.
Here $E_\mathrm{g}(\theta)$ is the energy of the self-consistent solution that is obtained under the constraint that
every Os moment is rotated from that of the ground state around the [001] axis, and its magnitude is equal to $\mos$ in the ground state.
One can clearly see that $E_\mathrm{g}$ remarkably increases with the rotation by as large as about 40 meV/Os and 70 meV/Os for $\Ueff=$ 1.25 and 2.0 eV, respectively.
This proves the existence of strong magnetic anisotropy in this compound even near the MIT. 

To gain deeper insight into the nature of the magnetic anisotropy,
we extend the analysis using a phenomenological model for the energy change associated with the rotations of the Os moments:
\begin{eqnarray}
	\mathcal{H}&=&J\sum_{\langle i,j \rangle, i<j}\vec{m}_i\cdot\vec{m}_j-\Asia\sum_i \left(\vec{m}_i\cdot \vec{\alpha}_i\right)^2\nonumber\\
	&& +\ADM\sum_{\langle i,j\rangle,i<j} \vec{d}_{ij} \cdot \left(\vec{m}_i\times\vec{m}_j\right),\label{eq:ham}
\end{eqnarray}
where $\vec{m}_i$ is a unit direction vector of the magnetic moment at Os site $i$.
Here $J$, $\Asia$, and $\ADM$ are the nearest-neighbor exchange interaction, the single-ion anisotropy, and the DM interaction, respectively.
Note that $\mos$ weakly depends on $\Ueff$ in the AFI phase, and this effect is renormalized into the interaction parameters.
The unit vector $\vec{\alpha}_i$ is along the local $\langle 111 \rangle$ axis at site $i$.
The case where $\Asia>0$ corresponds to the easy-axis anisotropy.
The unit vectors $\vec{d}_{ij}$ are the direction vectors of the DM interaction, which are the only one symmetry-allowed form~\cite{Elhajal05}\footnote{
The middle points of nearest-neighboring Os sites are not inversion centers of the pyrochlore lattice.
}.
In Ref.~\onlinecite{Elhajal05}, the cases where $\ADM>0$ and $\ADM<0$ are distinctly referred to as the ``direct case'' and ``indirect case'';
the former favors the all-in/all-out ordering.

The model parameters can be extracted by
fitting energies obtained in electronic structure calculations by Eq.~(\ref{eq:ham}).
Hereafter, the energy is always measured per Os atom.
In the following calculations, the magnitudes of the Os moments are fixed to $\mos$ of the ground state.
On the basis of Eq.~(\ref{eq:ham}), the energy difference between the all-in/all-out and 3-in/1-out states is given by
\begin{eqnarray}
	E_\mathrm{e}(0)-E_\mathrm{g}(0)&=&J+2\sqrt{2}\ADM.~\label{eq:flip}
\end{eqnarray}
Here, the 3-in/1-out state is the lowest excited state for $J>0$ and $\Asia>0$, which is obtained by flipping one of four non-equivalent spins in the all-in/all-out state [see Figs.~4(b) and (c)].
Changes in the energies of the all-in/all-out [$E_\mathrm{g}(\theta)$] and 3-in/1-out ordered states [$E_\mathrm{e}(\theta)$] are obtained respectively as follows:
\begin{eqnarray}
	E_\mathrm{g}(\theta)-E_\mathrm{g}(0)&=&E_\mathrm{e}(\theta)-E_\mathrm{e}(0)\nonumber \\
	&& +\frac{4\sqrt{2}}{3}\ADM\left(1-\cos\theta\right),\label{eq:E1} \\
	E_\mathrm{e}(\theta)-E_\mathrm{e}(0)&=&\Asia\left\{1-\frac{1}{9}\left(2\cos\theta+1\right)^2\right\},\label{eq:E2}
\end{eqnarray}
where $\theta$ denotes the rotation angle around the [001] axis.

Table~\ref{tbl:param} summarizes the model parameters extracted for $\Ueff=$ 1.25 and 2.0 eV.
The values of $\Asia$ and $\ADM$ were obtained by fitting the magnetic anisotropy energies at small $\theta$'s, i.e., in the range of $0^\circ \le \theta \le 60^\circ$
\footnote{The deviations of the data from the fitting curves for $\theta>60^\circ$ might come from higher order terms omitted in the simple model (\ref{eq:ham}).}.
We found that the values of $J$ and $\Asia$ are positive for both the abovementioned values of $\Ueff$'s.
The antiferromagnetic $J$ might originate in the superexchange coupling via the O site.
The easy-axis anisotropy $\Asia$ is found to be even larger than $J$ for both $\Ueff$'s.
On the other hand, $\ADM$ is considerably smaller than $\Asia$, and vanishes as $\Ueff$ increases.
This might be because the DM interaction appears as a perturbation with respect to hopping between the nearest-neighboring Os atoms.
These results suggest that the all-in/all-out order is mainly stabilized by the large easy-axis anisotropy rather than the DM interaction.
In the case of Y$_2$Ir$_2$O$_7$~\cite{Wan11}, the all-in/all-out ordering is ascribed to the direct DM interaction 
because Ir$^{4+}$ has an effective total angular momentum of $J_\mathrm{eff}=1/2$~\cite{Wan11,Kim08,Kim09}\footnote{No uniaxial single-ion anisotropy is present in $J_\mathrm{eff}=1/2$ or $S=1/2$ systems.}.
The observation of the large $\Asia$ indicates that Os$^{5+}$ has a larger value of $J_\mathrm{eff}~(J_\mathrm{eff}>1/2)$.

Assuming that the real material is located near the MIT, e.g, $\Ueff\simeq 1.25$ eV,
the calculated results provide a natural explanation for the peculiar low-$T$ properties of this compound.
(1) The large direct gap, $\Dd$, and small charge gap, $\Dc$, near the MIT are
consistent with the experimental observation of a large optical gap and the absence of a clear charge gap at low $T$'s.
In fact, $\Dd\simeq 0.17$ eV estimated at $\Ueff=1.25$ eV is comparable to the optical gap of about 0.1 eV observed below $\TMIT$~\cite{Padilla02}.
The \textit{pseudo gap} is expected to develop below $\TMIT$ concurrently with the onset of the N\'{e}el ordering.
The density of states vanishes toward low $T$'s in the wide energy range.
This observation explains the semiconducting behavior of resistivity up to the high $\TMIT$.
(2) The strong easy-axis anisotropy on the order of several tens meV stabilizes the all-in/all-out order cooperatively with the antiferromagnetic exchange interaction.
This can account for the high antiferromagnetic transition temperature of this compound.
The present result indicates that this compound has a magnetic gap of several tens meV.
Further experiments are needed to detect the magnetic gap and confirm the present observation.
\begin{figure}
 \centering
 \includegraphics[width=.475\textwidth,clip]{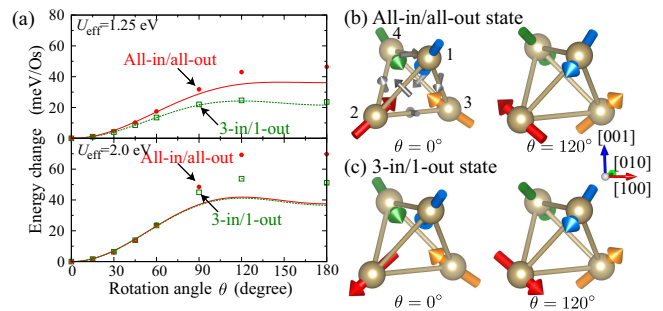}
 \caption{(color online).
 (a) Magnetic anisotropy energies estimated by the rotation of Os magnetic moments around the [001] axis.
(b)/(c) all-in/all-out and 3-in/1-out ordered states rotated by $\theta=0^\circ$ and $\theta=120^\circ$ around the [001] axis.
Every spin is perpendicular to the local $\langle 111 \rangle$ axis at $\theta=120^\circ$.
In (b), the small (gray) vectors and the labels represent the direction vector $\vec{d}_{ij}$ of the DM interaction and the site indices $i$, respectively.
}
 \label{fig:4}
\end{figure}

Before closing this Letter, we would like to comment on future work.
Contrary to the recent theoretical proposals~\cite{Guo09,Kargarian11}, 
all the three phases in the phase diagram have a trivial $Z_2$ topological invariant.
The band inversion in the AFM phase might suggest the presence of Dirac cones.
A detailed analysis of the nature of the band inversion is a part of our future study.
In contrast to Cd$_2$Os$_2$O$_7$, Hg$_2$Os$_2$O$_7$ remains metallic below the AF transition temperature~\cite{Reading02}.
This material might be located in the AFM phase.
The absence of a clear charge gap below the continuous MIT associated with the N\'{e}el ordering is also observed in NaOsO$_3$ perovskite~\cite{Shi09}.
Clearly, first-principle studies on this related compound will be interesting.

In summary, we performed LSDA+SO+$U$ calculations to explore the ground-state properties of Cd$_2$Os$_2$O$_7$.
We also found the all-in/all-out magnetic order as the ground state in a wide region of the phase diagram,
supporting recent X-ray experimental results.
We also showed that this magnetic order is stabilized by strong easy-axis anisotropy originating in spin-orbit coupling.
Furthermore, we found a \textit{pseudo gap} in the density of states near the metal-insulator transition.
These numerical results provide a natural explanation for the puzzling low-temperature properties of this compound.
The present result might open up new possibilities to explore Os compounds as a stimulating playground for the interplay of electron correlation and spin-orbit coupling.
\begin{table}
	\centering
	\begin{tabular}{lccc}
		\hline
		$\Ueff$ (eV) & $J$ (meV) & $\Asia$ (meV) & $\ADM$ (meV) \\
		\hline
		1.25 &   14 &  24 & 4 \\
		2.0  &   35 &  41 & 0 \\
		\hline
	\end{tabular}
	\caption{Estimated parameters for the phenomenological model given in Eq.~(\ref{eq:ham}).}
	\label{tbl:param}
\end{table}

We thank C. D. Batista, T. Kosugi, Z. Hiroi, Y. Motome, H. Ohnishi, T. Ozaki, M. Takigawa, K. Terakura, I. Yamauchi, and J. Yamaura for fruitful discussions.
Numerical calculations were partly carried out at the Supercomputer Center, ISSP, Univ. of Tokyo.
This work was supported by Grant-in-Aid for Scientific Research (No. 22104010), from MEXT, Japan.
A part of this research has been funded by the Strategic Programs for Innovative Research (SPIRE), MEXT, and by the Computational Materials Science Initiative (CMSI), Japan.


\end{document}